\documentclass{IET}

\renewcommand{\baselinestretch}{2}

\usepackage{amsmath,amssymb,amscd}
\usepackage{graphicx,cite}
\usepackage{algorithm}
\usepackage{algpseudocode}
\usepackage{subfigure}

\newcommand{\argmin}{\arg\!\min}
\newcommand{\argmax}{\arg\!\max}

\begin{document}
\renewcommand{\baselinestretch}{1}

\title{Positioning via Direct Localization in C-RAN Systems}

\author[1]{Seongah Jeong}
\affil{School of Engineering and Applied Sciences (SEAS), Harvard university, 29 Oxford street, Cambridge, MA 02138, USA}

\author[2]{Osvaldo Simeone}
\affil{Department of Electrical \& Computer Engineering, New Jersey Institute of Technology (NJIT),
Newark, NJ 07102, USA}

\author[2]{Alexander Haimovich}

\author[3,*]{Joonhyuk Kang}
\affil{Department of Electrical Engineering, Korea Advanced Institute of Science and Technology
(KAIST), 291, Daehak-ro, Yuseong-gu, Daejeon 34141, Korea}
\affil[*]{jhkang@ee.kaist.ac.kr}

\abstract{Cloud Radio Access Network (C-RAN) is a prominent architecture for 5G wireless cellular system that is based on the centralization of baseband processing for multiple distributed radio units (RUs) at a control unit (CU). In this work, it is proposed to leverage the C-RAN architecture to enable the implementation of \textit{direct localization} of the position of mobile devices from the received signals at distributed RUs. With ideal connections between the CU and the RUs, direct localization is known to outperform traditional \textit{indirect localization}, whereby the location of a source is estimated from intermediary parameters estimated at the RUs. However, in a C-RAN system with capacity limited fronthaul links, the advantage of direct localization may be offset by the distortion caused by the quantization of the received signal at the RUs. In this paper, the performance of direct localization is studied by accounting for the effect of fronthaul quantization with or without dithering. An approximate Maximum Likelihood (ML) localization is developed. Then, the Cram\'{e}r-Rao Bound (CRB) on the squared position error (SPE) of direct localization with quantized observations is derived. Finally, the performance of indirect localization and direct localization with or without dithering is compared via numerical results.}

\maketitle
\section{Introduction}\label{sec:intro}
Positioning of radio frequency (RF) sources by nodes of a wireless network is currently an active research area due to its relevance in Global Positioning System (GPS)-denied environments, such as military applications, indoors localization, disaster response, emergency relief, surveillance and tactical systems [1]-[12]. The importance of the topic is attested by the recent Federal Communications Commission (FCC) specification for indoor positioning accuracy [13] and by the inclusion of various localization techniques in the Long Term Evolution (LTE) standard [14]. The traditional approach to source localization has been \textit{indirect, or two-step, localization}. Accordingly, a set of distributed base stations (BSs) estimates position-related parameters, such as time of arrival (TOA) [1]-[5], [7], time difference of arrival (TDOA) [1], [3], angle of arrival (AOA) [1], [4], [6]-[11] or received signal strength (RSS) [4], [12]. The estimated parameters are then transmitted to a control unit (CU) that determines the source's position. For instance, in LTE, positioning is based on TDOA in the uplink, which is referred to as uplink TDOA (UTDOA) [14].   
 
Cloud radio access network (C-RAN), a cloud-based network architecture, has emerged as a leading technology for 5G and beyond [15]. In C-RANs, BSs are connected to a CU by means of so called fronthaul links, as shown in Fig. \ref{fig:sys}. The fronthaul links carry sampled and quantized baseband signals to and from the CU. Thanks to the fronthaul links, the CU can carry out all baseband processing on behalf of the connected BSs. Since the BSs implement only radio functionalities, they are typically referred to as radio units (RUs). C-RANs hence enable the joint processing at the CU of the signals received by the RUs, making it possible to implement \textit{direct, or one-step, localization} [1], [3], [7]-[11], whereby the source's position is determined directly from the received signals at the RUs. The concept of direct localization was first introduced by Wax and Kailath [10] in the 1980’s. However, it was only recently that efficient algorithms were proposed for its implementation [1].   

With ideal fronthaul links, direct localization can generally outperform indirect localization [1], [7]-[9]. However, in a C-RAN, due to the limited capacity of the fronthaul links, the baseband signals received by the RUs are quantized prior to transmission to the CU. In this work, we study the performance of direct localization by accounting for the distortion caused by fronthaul quantization. While in [1], [7]-[11], direct localization in C-RAN systems was studied by assuming ideal compression within an information-theoretic framework, here we consider a more practical scenario in which the RUs perform scalar quantization as in state-of-the-art C-RAN systems. 

In particular, in considered C-RAN systems, fronthaul transmission is based on scalar quantization that allows for an arbitrary additive dither. Dithering, that is the addition of a random offset, known at the receiver, prior to quantization, is known to provide potential gains over conventional quantization [16], [17]. In the context of distributed detection, as for the C-RAN systems at hand, the performance gain of dithering can be ascribed to the capability to effectively enhance the resolution of the observation at the CU. To see this, consider the case in which each RU has a one-bit threshold quantizer per quadrature component. The use of independent random dithering at the RUs allows each RU to effectively use a different, and randomly selected, threshold for quantization, hence improving the resolution of the effective quantizer between the RUs and the CU.

The main contributions of this paper are summarized as follows.
\begin{itemize}
\item  To the best of our knowledge, this is the first work in the literature, except our prior work [3], to analyze direct localization in C-RAN with limited capacity links. Unlike the information-theoretic analysis of [3], here we assume practical fronthaul processing based on scalar quantization, enabling also dithering. 
\item We derive an approximate Maximum Likelihood (ML) solution for direct localization in C-RAN system.       
\item We derive the Cram\'{e}r-Rao Bound (CRB) on the accuracy of direct localization by taking into account the impact of quantization. The analysis allows us to analytically quantify the performance loss caused by the fronthaul quantization. We also show that for asymptotically large fronthaul capacity, that CRB tends to that of direct localization from unquantized received signals.      
\item The performances of the proposed direct localization schemes with and without dithering are compared with direct and indirect localization via ideal fronthaul links as well as with the relevant CRBs in C-RAN systems, via numerical results.  
\end{itemize}

In the following, after introducing the system and signal models, we shortly overview indirect localization in Section \ref{sec:SysSigPre}. Novel direct localization schemes based on approximate ML are proposed for the C-RAN system in Section \ref{sec:Di}, and the analytical performance of direct localization via CRB is derived in Section \ref{sec:CRB}. Numerical examples and concluding remarks are presented in Section \ref{sec:num} and Section \ref{sec:con}, respectively.
\section{System, Signal Models and Preliminaries}\label{sec:SysSigPre}
In this section, we describe system and signal models, and review indirect localization.
\subsection{System Model}\label{sec:sys}
\begin{figure}[t]
\begin{center}
\includegraphics[width=8.5cm]{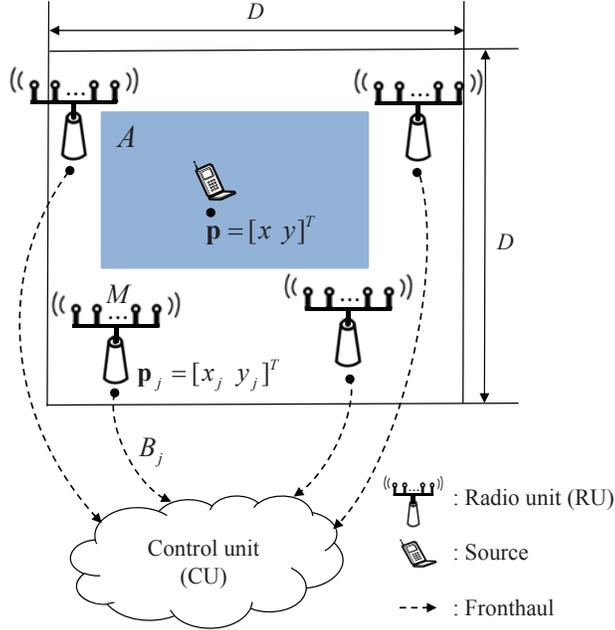}
\caption{Illustration of the considered C-RAN system used for localization.} \label{fig:sys}
\end{center}
\end{figure}

We consider the C-RAN system illustrated in Fig. \ref{fig:sys}, which consists of a single active source, $N_r$ distributed RUs and a CU. The RUs may consist of different types of infrastructure nodes such as macro/femto/pico base stations, relay stations or distributed antennas, and the source may be a mobile device. Each RU is equipped with an $M$-element antenna array, while the source is a single-antenna node. The set of RUs, denoted as $\mathcal{N}_r =\{1, \dots, N_r\}$, is located within a $D \times D$ square area, and all the RUs are assumed to be synchronous (e.g., via GPS). The source is located at a position $\pmb{p}=[x\,\,y]^T$, which is known a priori to lie in a given region $A$ contained within overall square region. Each RU $j \in \mathcal{N}_r$ is located at a position $\pmb{p}_j=[x_j\,\,y_j]^T$, and the positions of all RUs are assumed to be known to the CU. The distance and angle between the source and RU $j$ are defined as 
\vspace{-1cm}
\begin{subequations}
\begin{eqnarray}
d_j(\pmb{p}) &=& \|\pmb{p}-\pmb{p}_j\|, \\
\phi_j(\pmb{p}) &=& \tan^{-1}\left(\frac{y-y_j}{x-x_j}\right), 
\end{eqnarray}
\end{subequations}
respectively. We assume far-field conditions, namely that the array aperture is much smaller than the distance $d_j(\pmb{p})$.

The RUs are connected to the CU via finite-capacity fronthaul links, and the CU aims at locating the source based on quantized data transmitted from the RUs to the CU over the fronthaul links. Specifically, each RU $j$ sends a message at a rate of $B_j$ bps/Hz to the CU, where the normalization is with respect to the bandwidth of the signal transmitted by the source, as we will further discuss below. Note that the fronthaul links can be either wireless, e.g., a microwave link, or wired, e.g., a coaxial cable or a fiber optics link.
\subsection{Signal Model}\label{sec:sig}
The source transmits a scaled version of a baseband waveform $s(t)$ with unit energy and single sided bandwidth $W$ Hz at an unknown time $t_0$. The waveform $s(t)$ is known to the CU and it may represent a training or synchronization signal. During the observation time $T$, the $M \times 1$ received signal vector at the RU $j \in \mathcal{N}_r$ is hence expressed as
\begin{equation}\label{eq:RUsig}
\pmb{r}_j(t)=b_j\pmb{\alpha}_j(\pmb{p})s(t-\tau_j(\pmb{p})-t_0)+\pmb{z}_j(t), \hspace{0.3cm} 0 \le t \le T,
\end{equation}
where $b_j$ is an unknown complex scalar coefficient whose average power $E[|b_j|^2]$ is only available at the CU and that describes the channel attenuation from the source to RU $j$; $\pmb{\alpha}_j(\pmb{p})$ is the antenna array response of RU $j$, which, assuming a uniform linear antenna array (ULA) at each RU, is given by $\pmb{\alpha}_j(\pmb{p})=1/\sqrt{M}[1 \,\,\, e^{-i2\pi \Delta_A\cos\phi_j(\pmb{p})/\lambda} \cdots $ $e^{-i2\pi(M-1) \Delta_A\cos\phi_j(\pmb{p})/\lambda}]^T$, with the wavelength of the signal $\lambda$ and the antenna separation $\Delta_A$; $\tau_j(\pmb{p})$ is the propagation delay along the path between the source and RU $j$, which depends on the source's position as $\tau_j(\pmb{p})=d_j(\pmb{p})/c$, with the propagation speed $c$; and $\pmb{z}_j(t)$ represents the contribution of noise and interference whose components are assumed to a zero-mean,  white, complex Gaussian stochastic process with power $\sigma_j^2$. We assume that the source moves slowly enough and that the observation time $T$ is sufficiently long so that the inequality $T \gg \max_{j \in \mathcal{N}_r}{\tau_j(\pmb{p})}+t_0$ is satisfied. 

After sampling at Nyquist rate, the discrete-time received signal (\ref{eq:RUsig}) is given as
\begin{equation}\label{eq:RUsample}
\pmb{r}_j[n]\triangleq \pmb{r}_j(nT_s)=b_j\pmb{\alpha}_j(\pmb{p})s_j[n]+\pmb{z}_j[n], \hspace{0.5cm}0 \le n \le N_s-1,\\
\end{equation}
where $T_s=1/2W$ is the sampling period; $N_s=\lceil T/T_s \rceil$ is the number of time samples; and the sampled versions of the signal and the noise are denoted $s_j[n] \triangleq s(nT_s-\tau_j(\pmb{p})-t_0)$ and $\pmb{z}_j[n] \triangleq \pmb{z}_j(nT_s)$, respectively. Taking the discrete Fourier transform (DFT) of the discrete signal (\ref{eq:RUsample}), the received signal is expressed in the frequency domain as 
\begin{equation}\label{eq:RUdft}
\pmb{R}_j(k)=b_j\pmb{\alpha}_j(\pmb{p})S(k)e^{-iw_k\left(\tau_j(\pmb{p})+t_0\right)}+\pmb{Z}_j(k), \hspace{0.5cm}0 \le k \le N_s-1,\\
\end{equation} 
where $w_k=2\pi k/(N_sT_s)$; and $\pmb{R}_j(k)$, $S(k)$ and $\pmb{Z}_j(k)$ indicate the DFT coefficients of the respective time-domain signals $\pmb{r}_j[n]$, $s(nT_s)$, and $\pmb{z}_j[n]$. Given the assumptions above, we have the equality $\sum_{k=0}^{N_s-1}|S(k)|^2 = 1$, and the DFT coefficients of the noise $\pmb{Z}_j(k)$ are uncorrelated and have the power $\sigma_j^2$, for $0 \le k \le N_s-1$. We observe that, in (\ref{eq:RUdft}), the information about the source's location is embedded in two ways in the received signal, namely in time delay $\tau_j(\pmb{p})$ and in the array response $\pmb{\alpha}_j(\pmb{p})$.

The RUs quantize and forward the received signals in (\ref{eq:RUdft}) to the CU for positioning under the fronthaul constraint.  As a practical solution, prescribed, e.g., by the CPRI standard [19], we assume that each RU applies the same uniform quantizer for all frequencies and antennas. To evaluate the localization accuracy, we adopt the SPE [2], [3], [18]
\begin{equation}\label{eq:spe}
\rho=\mathrm{E}\left[\left\|\hat{\pmb{p}}-\pmb{p}\right\|^2\right],
\end{equation}
as the performance criterion of interest, where $\hat{\pmb{p}}$ is the estimated source's position. 
\subsection{Indirect Localization in C-RAN}\label{sec:InDi}
In the conventional, indirect localization method, each RU $j$ estimates the parameters $\tau_j(\pmb{p})+t_0$ and $\phi_j(\pmb{p})$, which are referred to as TOA and AOA measurements, respectively. Subsequently, these estimates are sent to the CU, which locates the source, for example, by maximum likelihood based on TOA and AOA estimates, where the dependence on the unknown parameter $t_0$ can be eliminated. Note that, under fronthaul capacity constraints, the TOA and AOA measurements at RU $j$ should be quantized prior to transmission to the CU. Nevertheless, assuming that the number of samples $N_s$ is large enough, the number of bits $N_sB_j$ available for quantization of the estimates is large enough to make the corresponding distortion due to quantization negligible.

\section{Direct Localization in C-RAN}\label{sec:Di}
Here, we derive an approximate ML solution for direct localization in C-RAN system with limited fronthaul links, for which a practical fronthaul processing is considered based on scalar quantization with dithering. The approximation is taken so as to account for the effect of scalar quantization and for the optimal solution of the likelihood function, whose details are discussed as follows.     

With direct localization, the position of the source is estimated directly from the signals at the RUs [1], [3], [7]-[9]. Due to the fronthaul capacity limitation of the C-RAN system, we assume that each RU $j$ sends the quantized version of the received signal $\pmb{R}_j(k)$ in (\ref{eq:RUdft}) to the CU over limited-capacity fronthaul links with $B_j$ bits per sample. The CU performs localization based on the quantized signals. Specifically, as discussed below, we adopt a standard uniform dithered quantization [16], [17], where the dither is known to the CU. The addition of a dither signal is known to potentially improve the accuracy of a quantized signal as discussed in Section \ref{sec:intro}.   

Scalar quantization at RU $j$ operates on the dithered signal 
\begin{equation}\label{eq:Dsig}
\pmb{R}_j(k)+\pmb{D}_j(k),
\end{equation}
where $\pmb{D}_j(k)$ is a dither signal, independent of the received signal $\pmb{R}_j(k)$. Quantization uses $B_j/2M$ bits, or equivalently $L_j=2^{B_j/2M}$ quantization levels, for either the real or imaginary parts of the dithered signal (\ref{eq:Dsig}) at any of the $M$ received antennas. The dynamic range $[-R_j^{\max}, R_j^{\max}]$ of the quantizer applied at each RU $j$ is fixed, and may be based on preliminary Monte Carlo experiments aimed at guaranteeing that the peak-to-peak interval includes the received signals $\{\pmb{R}_j(k)\}$ with high probability, when considering a given random distribution of the source, and of the channel coefficients, as well as the distribution of the noise (see also Section \ref{sec:num}). Each real and imaginary entry of the dithered signal (\ref{eq:Dsig}) is quantized to the closest value $\hat{\pmb{R}}_j(k)$ in the set of representation points $\{-R_j^{\max}+(l-1)\Delta_j\}$ for $l=1, \dots, L_j$, where $\Delta_j=2R_j^{\max}/(L_j-1)$ is the step size of the quantizer. We take the dithers $\pmb{D}_j(k)$ to be independent and uniformly distributed in the interval $[-\Delta_j/I_j, \Delta_j/I_j]$, where $I_j$ is a parameter that can be chosen to minimize the SPE (\ref{eq:spe}).  

The CU recovers the quantized versions $\{\hat{\pmb{R}}_j(k)\}$ of the dithered received signal, i.e., of $\{\pmb{R}_j(k)+\pmb{D}_j(k)\}$, from all RU $j \in \mathcal{N}_r$. Then, it subtracts the dither to obtain 
\begin{equation}\label{eq:Subsig}
\tilde{\pmb{R}}_j(k)=\hat{\pmb{R}}_j(k)-\pmb{D}_j(k).
\end{equation}
Differently from [1], [7]-[9], the estimator under consideration is applied to the signals $\{\tilde{\pmb{R}}_j(k)\}$ in (\ref{eq:Subsig}), rather than to the received signals $\{\pmb{R}_j(k)\}$ due to the limited fronthaul capacity. We are interested in approximating the ML estimate of the position $\pmb{p}$ based on the received signal $\{\tilde{\pmb{R}}_j(k)\}$ in (\ref{eq:Subsig}). 

To this end, we account for the effect of quantization as an additive Gaussian noise term, as in 
\begin{equation}\label{eq:Qsig}
\tilde{\pmb{R}}_j(k)=\pmb{R}_j(k)+\pmb{Q}_j(k),
\end{equation}
where $\pmb{Q}_j(k)$ represents the quantization noise whose components have the zero-mean Gaussian distribution with the variance $\sigma_{q,j}^2(k)$. In this way, from (\ref{eq:RUdft}), the effective Gaussian noise on the observation $\{\tilde{\pmb{R}}_j(k)\}$ of RU $j$ is given by $\pmb{Z}_j(k)+\pmb{Q}_j(k)$, whose entries are distributed i.i.d. as zero-mean complex Gaussian variables with variance $\gamma_j^2(k)=\sigma_j^2(k)+\sigma_{q,j}^2(k)$.   

With reference to (\ref{eq:Subsig}), an approximate ML estimate of the unknown parameters $(\{b_j\}, t_0, \pmb{p})$ is hence obtained by solving the problem
\begin{equation}\label{eq:BossCost}
\text{minimize} \,\,\, \sum_{j=1}^{N_r}C_j(b_j, t_0, \pmb{p}),
\end{equation}
with
\begin{equation}\label{eq:Dcost}
C_j(b_j, t_0, \pmb{p})=
\sum_{k=0}^{N_s-1}\frac{1}{\gamma_j^2(k)}\left\|\tilde{\pmb{R}}_j(k)-b_j\pmb{\alpha}_j(\pmb{p})S(k)e^{-iw_k\left(\tau_j(\pmb{p})+t_0\right)}\right\|^2
\end{equation}
over the position $\pmb{p}$, the transmit time $t_0$ and the channel attenuation parameters $\{b_j\}$. 

To evaluate the power of the quantization noise $\sigma_{q,j}^2(k)$, and hence $\gamma_j^2(k)=\sigma_j^2(k)+\sigma_{q,j}^2(k)$,  as a function of the fronthaul capacity $B_j$, we leverage arguments from rate-distortion theory [20]. Note that this provides an approximation of the effect of quantization since scalar quantization is adopted here, as opposed to the vector quantization with arbitrarily large blocks that is assumed in rate-distortion theory. This approximation allows us to compute $\sigma_{q,j}^2(k)$ as the value that guarantees the equality [21, eq. (10.11)] 
\begin{equation}\label{eq:mutual}
I([\pmb{R}_j(k)]_m; [\tilde{\pmb{R}}_j(k)]_m)=\frac{B_j}{M},
\end{equation}
where $I([\pmb{R}_j(k)]_m; [\tilde{\pmb{R}}_j(k)]_m)$ is the mutual information between the input of the quantizer and its output, after subtraction of the dither. In order to evaluate (\ref{eq:mutual}), we take $[\pmb{R}_j(k)]_m$ to be distributed as $\mathcal{CN}(0, E[|b_j|^2]|S(k)|^2+\sigma_j^2)$, hence treating the parameter $b_j$ as complex Gaussian with zero mean and power $E[|b_j|^2]$, i.e., Rayleigh fading. The net effect of these choices is that the weights $1/\gamma_j^2(k)$ in (\ref{eq:Dcost}) are given by  
\begin{equation}\label{eq:weight}
\gamma_j^2(k)= \sigma_j^2+\frac{E[\left|[\pmb{R}_j(k)]_m\right|^2]}{2^{\frac{B_j}{M}}-1}
= \sigma_j^2+\frac{E[|b_j|^2]|S(k)|^2+\sigma_j^2}{2^{\frac{B_j}{M}}-1}.
\end{equation}
As per (\ref{eq:weight}), the weight given to each RU $j$ in (\ref{eq:Dcost}) increases with the capacity $B_j$ of its fronthaul link. 

The minimization (\ref{eq:BossCost}) is first solved over the amplitude parameter $b_j$ for fixed $(t_0, \pmb{p})$, yielding
\begin{eqnarray}\label{eq:bopt}
b_j^*(t_0, \pmb{p})&=&\argmin_{b_j}\sum_{j=1}^{N_r}C_j(b_j, t_0, \pmb{p})\nonumber\\
&=&\frac{1}{\left\|\bar{\pmb{S}}_j^\gamma(t_0, \pmb{p})\right\|^2\left\|\pmb{\alpha}_j(\pmb{p})\right\|^2}\left[\bar{\pmb{S}}_j^\gamma(t_0, \pmb{p})\otimes\pmb{\alpha}_j(\pmb{p})\right]^H\tilde{\pmb{R}}_j^\gamma,\nonumber\\
&=&\frac{1}{\left\|\bar{\pmb{S}}_j^\gamma(t_0, \pmb{p})\right\|^2}\left[\bar{\pmb{S}}_j^\gamma(t_0, \pmb{p})\otimes\pmb{\alpha}_j(\pmb{p})\right]^H\tilde{\pmb{R}}_j^\gamma,
\end{eqnarray}
where $\bar{\pmb{S}}_j^\gamma(t_0, \pmb{p})=[S(0)e^{-iw_0(\tau_j(\pmb{p})+t_0)}/\gamma_j(0) \cdots S(N_s-1)e^{-iw_{N_s-1}(\tau_j(\pmb{p})+t_0)}/\gamma_j(N_s-1)]^T$; $\tilde{\pmb{R}}_j^\gamma=[\tilde{\pmb{R}}_j^{T}(0)/\gamma_j(0) \cdots \tilde{\pmb{R}}_j^{T}(N_s-1)/\gamma_j(N_s-1)]^T$; and the last equality can be obtained by recalling that we have $\left\|\pmb{\alpha}_j(\pmb{p})\right\|^2=1$ for $j \in \mathcal{N}_r$. By substituting (\ref{eq:bopt}) in (\ref{eq:Dcost}) and simplifying the notation as $\|\bar{\pmb{S}}_j^\gamma\|^2=\|\bar{\pmb{S}}_j^\gamma(t_0, \pmb{p})\|^2=\sum_{k=0}^{N_s-1}|S(k)|^2/\gamma_j^2(k)$, we obtain that the minimization (\ref{eq:BossCost}) is equivalent to the problem
\begin{equation}\label{eq:Max1}
\text{maximize} \,\,\, \tilde{C}(t_0, \pmb{p}),
\end{equation}
with 
\begin{subequations}\label{eq:DcostRe}
\begin{eqnarray}
\tilde{C}(t_0, \pmb{p}) &=& \sum_{j=1}^{N_r} \left| \frac{\pmb{\alpha}_j^H(\pmb{p})}{\left\|\bar{\pmb{S}}_j^\gamma\right\|} \sum_{k=0}^{N_s-1} \frac{e^{iw_k\left(\tau_j(\pmb{p})+t_0\right)}S^*(k)}{\gamma_j^2(k)}\tilde{\pmb{R}}(k)\right|^2\\ 
&=&\left\|\pmb{z}^H(t_0)\pmb{S}^H\pmb{U}(\pmb{p})\right\|^2\\
&=&\sum_{j=1}^{N_r}\left|\pmb{z}^H(t_0)\pmb{V}_j(\pmb{p})\right|^2, \label{eq:fft}
\end{eqnarray}
\end{subequations}
where $\pmb{z}(t_0)=[1 \,\, e^{-iw_1t_0} \cdots e^{-iw_{N_s-1}t_0}]^T$;  $\pmb{S}=\text{diag}\{S(0), \dots, S(N_s-1)\}$; and $\pmb{U}(\pmb{p})=[\pmb{u}_1(\pmb{p})/\|\bar{\pmb{S}}_1^\gamma\|$ $\cdots \pmb{u}_{N_r}(\pmb{p})/\|\bar{\pmb{S}}_{N_r}^\gamma\|]$, with $\pmb{u}_j(\pmb{p})=[e^{iw_0\tau_j(\pmb{p})}$ $\pmb{\alpha}_j^H(\pmb{p})\tilde{\pmb{R}}_j(0)/\gamma_j^2(0) \,\,\,\,\, \cdots \,\,\,\,\, e^{iw_{N_s-1}\tau_j(\pmb{p})}\pmb{\alpha}_j^H(\pmb{p})\tilde{\pmb{R}}_j(N_s-1)$ $/\gamma_j^2(N_s-1)]^T$; and $\pmb{V}_j(\pmb{p})$ is the $j$th column of $\pmb{V}(\pmb{p})=\pmb{S}^H\pmb{U}(\pmb{p})$.

Since the vector $\pmb{z}(t_0)$ in (\ref{eq:fft}) has the structure of a column of a $N_s$-point Discrete Fourier Transform (DFT), the evaluation of (\ref{eq:fft}) for different potential value of $t_0$ can be computed by means of Fast Fourier Transform (FFT), allowing us to efficiently estimate $t_0$ in the following way [1], [7]. Define a grid of possible values for $t_0$ as $\{0, 1, \dots, q_{t_0}N_s-1\}T_s/(q_{t_0}N_s)$, where $q_{t_0}$ is an integer that determines the resolution of the estimate of $t_0$. Apply the $q_{t_0}N_s$-point FFT matrix $\pmb{F}$ to each column $\pmb{V}_j(\pmb{p})$ for $j \in \mathcal{N}_r$, where $[\pmb{F}]_{k,n}=e^{-i2\pi (k-1)(n-1)/(q_{t_0}N_s)}$ for $1 \le k,n \le q_{t_0}N_s$, with $[\pmb{A}]_{k,n}$ defining the $(k,n)$th element of matrix $\pmb{A}$. Note that, when $q_{t_0}\ge 2$, each vector $\pmb{V}_j(\pmb{p})$ is padded with trailing zeros to length $q_{t_0}N_s$. Then, the maximization (\ref{eq:Max1}) can be carried out jointly over $t_0$ and $\pmb{p}$, yielding the position estimate $\pmb{p}^*$, by solving the following problem  
\begin{equation}\label{eq:FinalCost}
\pmb{p}^*=\argmax_{\pmb{p}}\left\{\max_{1 \le k \le q_{t_0}N_s} \left[\sum_{j=1}^{N_r}\left|\pmb{F}\pmb{V}_j(\pmb{p})\right|^2\right]_k \right\},
\end{equation}
where $[\pmb{a}]_{k}$ is the $k$th element of vector $\pmb{a}$; and $|\cdot|$ represents the entry-wise absolute value of its argument.  

In summary, the proposed solution is referred to an approximate ML because $(i)$ the effect of scalar quantization in approximated by a Gaussian noise term in (\ref{eq:Qsig}), whose variance is obtained by means of rate-distortion arguments; $(ii)$ the optimization over $t_0$ is obtained by means of the grid search described above.  
\section{CRB for Direct Localization}\label{sec:CRB}
In this section, we derive the CRB [18] on the SPE (\ref{eq:spe}) for the direct localization. The CRB in the presence of dithering generally depends on the distribution of the dither and its calculation appears to be intractable. Nevertheless, in the high SNR regime, where the CRB becomes tight, the impact of dithering is less pronounced. For these reasons, we concentrate here on the derivation of the CRB for standard quantization without dithering. The CRB provides a lower bound on the SPE (\ref{eq:spe}) as
\begin{equation}\label{eq:crb}
\rho \ge \text{tr}\left\{\pmb{J}^{-1}(\pmb{p})\right\},
\end{equation}
where $\pmb{J}(\pmb{p})$ is the Equivalent Fisher Information Matrix (EFIM) [2], [3] for the estimation of the source's position $\pmb{p}$.

To elaborate, we denote the uniform quantization function at RU $j$ as $Q_j(x)=\{l| q_{j,l-1}(\Delta_j) < x \le q_{j,l}(\Delta_j);\,\, l=\{1, \dots,L_j\} \}$, where $q_{j,l}(\Delta_j)=-R_j^{\max}+(l-0.5)\Delta_j$ is the $l$th quantization threshold for the RU $j$ with $q_{j,0}=-\infty$ and $q_{j,L_j}=\infty$. The signal $\pmb{S}=\{S(k)\}$ is known to the CU, and we also assume that the transmit time $t_0$ is known in order to obtain a lower bound on the SPE. The probability of the quantized signals available at the CU for a given unknown parameter vector $\pmb{\theta}=[\pmb{p}^T \,\, \pmb{b}_1^T \cdots \pmb{b}_{N_r}^T]^T$ with $\pmb{b}_j=[b_j^{\Re}\,\,b_j^{\Im}]^T$ is then given as
\begin{equation}\label{eq:cond}
P(\hat{\pmb{R}};\pmb{\theta})=\prod_{j=1}^{N_r}\prod_{k=0}^{N_s-1}\prod_{m=1}^{M}\prod_{\zeta \in \{\Re, \Im\}}\prod_{l=1}^{L_j}P_{j,l}(Q_j([\pmb{R}_j^\zeta(k)]_m); \pmb{\theta}_j)^{\delta\left(Q_j([\pmb{R}_j^\zeta(k)]_m)-l\right)},
\end{equation}
where $\hat{\pmb{R}}=[\hat{\pmb{R}}_1^T \cdots \hat{\pmb{R}}_{N_r}^T]^T$ with $\hat{\pmb{R}}_j=[\hat{\pmb{R}}_j^T(0) \cdots \hat{\pmb{R}}_j^T(N_s-1)]^T$; $\delta(\cdot)$ is the Kronecker-delta function, i.e., $\delta(0)=1$ and $\delta(x)=0$ for $x \neq 0$; and $P_{j,l}$ $(Q_j([\pmb{R}_j^\zeta(k)]_m); \pmb{\theta}_j)$ is the probability that either the real or imaginary part of the received signal, $[\pmb{R}_j^\Re(k)]_m$ or $[\pmb{R}_j^\Im(k)]_m$, for the $m$th antenna at the RU $j$ takes on a specific value $l$ among the $L_j$ quantization levels, i.e., $Q_j([\pmb{R}_j^\zeta(k)]_m)=l$, given as   
\begin{equation}\label{eq:prob}
P_{j,l}(Q_j([\pmb{R}_j^\zeta(k)]_m); \pmb{\theta}_j)=\Phi\left(\frac{q_{j,l}(\Delta_j)-f_{j,k,m}^\zeta(\pmb{\theta}_j)}{\sigma_j/\sqrt{2}}\right)-\Phi\left(\frac{q_{j,l-1}(\Delta_j)-f_{j,k,m}^\zeta(\pmb{\theta}_j)}{\sigma_j/\sqrt{2}}\right),
\end{equation}
with $\zeta$ being the parameter to indicate the real and imaginary part of the signal as $\zeta \in \{\Re, \Im\}$ and $\pmb{\theta}_j=[\pmb{p}^T\,\,\pmb{b}_j^T]^T$. In (\ref{eq:prob}), $\Phi(\cdot)$ is the complementary cumulative distribution function of the standard Gaussian distribution $\Phi(x)=\int_{-\infty}^x e^{-t^2/2}/\sqrt{2\pi}dt$, and we have defined the noiseless received signal (see (\ref{eq:RUdft})) as $f_{j,k,m}(\pmb{\theta}_j)= b_j[\pmb{\alpha}_j(\pmb{p})]_mS(k)e^{-iw_k(\tau_j(\pmb{p})+t_0)}$. 

The log-likelihood function is given by $L(\pmb{\theta})=\ln P(\hat{\pmb{R}}; \pmb{\theta})$, and the FIM $\pmb{J}(\pmb{\theta})$ for the unknown parameter vector $\pmb{\theta}$ can be written as $\pmb{J}(\pmb{\theta})=-E_{\hat{\pmb{R}}}[\nabla_{\pmb{\theta}}\nabla_{\pmb{\theta}}^TL(\pmb{\theta})]$ [18]. The CRB in (\ref{eq:crb}) can be then obtained from the EFIM for the location $\pmb{p}$, which is given as
\begin{equation}\label{eq:efim}
\pmb{J}(\pmb{p})=\pmb{X}-\pmb{Y}\pmb{Z}^{-1}\pmb{Y}^T,
\end{equation}
where $\pmb{X}=[\pmb{J}(\pmb{\theta})]_{(1:2,1:2)}$; $\pmb{Y}=[\pmb{J}(\pmb{\theta})]_{(1:2,3:2N_r+2)}$; and $\pmb{Z}=[\pmb{J}(\pmb{\theta})]_{(3:2N_r+2,3:2N_r+2)}$, with $[\pmb{A}]_{(a:b,c:d)}$ being the sub-matrix of $\pmb{A}$ corresponding to from the $a$th to the $b$th rows and from the $c$th to the $d$th columns. In Appendix \ref{app:efim}, we calculate these matrices as $\pmb{X}=\sum_{j=1}^{N_r}\pmb{U}_j\pmb{\Psi}_j\pmb{U}_j^T$; $\pmb{Y}=[\pmb{U}_1\pmb{\Psi}_1\pmb{V}^T \cdots \pmb{U}_{N_r}\pmb{\Psi}_{N_r}\pmb{V}^T]$; and $\pmb{Z}=\text{diag}\{\pmb{V}\pmb{\Psi}_1\pmb{V}^T, \dots, \pmb{V}\pmb{\Psi}_{N_r}\pmb{V}^T\}$, where $\pmb{V} = [\pmb{0}_2 \,\,\pmb{0}_2 \,\,\pmb{I}_2]$; $\pmb{U}_j = [\cos \phi_j(\pmb{p})/c \,\,\, -\sin \phi_j(\pmb{p})/d_j(\pmb{p}) \,\,\, \pmb{0}_2^T;$ $\sin \phi_j(\pmb{p})/c\,\,\, \cos \phi_j(\pmb{p})/d_j(\pmb{p})\,\,\, \pmb{0}_2^T]$; and the matrix $\pmb{\Psi}_j$ represents the FIM of RU $j$ with respect to $\tilde{\pmb{\theta}}_j= [\tau_j(\pmb{p})\,\,\phi_j(\pmb{p})\,\,\pmb{b}_j^T]^T$ and is defined as  
\begin{equation}\label{eq:psi}
[{\pmb{\Psi}_j}]_{p,q}=\sum_{k,m,\zeta,l}\frac{\left(\Gamma_{j,k,m,l}^\zeta-\Gamma_{j,k,m,l-1}^\zeta\right)^2\nabla_{[\tilde{\pmb{\theta}}_j]_p}f_{j,k,m}^\zeta(\pmb{\theta}_j)\nabla_{[\tilde{\pmb{\theta}}_j]_q}^Tf_{j,k,m}^\zeta(\pmb{\theta}_j)}{\pi\sigma_j^2 P_{j,l}(Q_j([\pmb{R}_j^\zeta(k)]_m); \pmb{\theta}_j)},
\end{equation}
for $1 \le p,q \le 4$, with being zero if the corresponding parameters are orthogonal,
\begin{equation}
\Gamma_{j,k,m,l}^\zeta = e^{-\frac{\left(q_{j,l}(\Delta_j)-f_{j,k,m}^\zeta(\pmb{\theta}_j)\right)^2}{\sigma_j^2}},
\end{equation}
and the required derivatives 
\begin{eqnarray}
\nabla_{\tau_j(\pmb{p})} f_{j,k,m}^\Re(\pmb{\theta}_j) &=& \mathrm{Re}\{-iw_k f_{j,k,m}(\pmb{\theta}_j)\}\nonumber\\
\nabla_{\tau_j(\pmb{p})}f_{j,k,m}^\Im(\pmb{\theta}_j) &=& \mathrm{Im}\{-iw_k f_{j,k,m}(\pmb{\theta}_j)\}\nonumber\\
\nabla_{\phi_j(\pmb{p})} f_{j,k,m}^\Re(\pmb{\theta}_j) &=& \mathrm{Re}\left\{\frac{i2\pi\Delta_A\sin \phi_j(\pmb{p})(m-1)f_{j,k,m}(\pmb{\theta}_j)}{\lambda}\right\}\nonumber\\
\nabla_{\phi_j(\pmb{p})} f_{j,k,m}^\Im(\pmb{\theta}_j) &=& \mathrm{Im}\left\{\frac{i2\pi\Delta_A\sin \phi_j(\pmb{p})(m-1)f_{j,k,m}(\pmb{\theta}_j)}{\lambda}\right\}\nonumber\\
\nabla_{b_j^\Re}f_{j,k,m}^\Re(\pmb{\theta}_j) &=& \mathrm{Re}\{[\alpha_j(\pmb{p})]_mS(k)e^{-iw_k(\tau_j(\pmb{p}+t_0))}\}\nonumber\\
\nabla_{b_j^\Re} f_{j,k,m}^\Im(\pmb{\theta}_j) &=& \mathrm{Im}\{[\alpha_j(\pmb{p})]_mS(k)e^{-iw_k(\tau_j(\pmb{p}+t_0))}\}\nonumber\\
\nabla_{b_j^\Im} f_{j,k,m}^\Re(\pmb{\theta}_j) &=& -\mathrm{Im}\{[\alpha_j(\pmb{p})]_m(k)e^{-iw_k(\tau_j(\pmb{p}+t_0))}\}\nonumber\\
\nabla_{b_j^\Im} f_{j,k,m}^\Im(\pmb{\theta}_j) &=& \mathrm{Re}\{[\alpha_j(\pmb{p})]_mS(k)e^{-iw_k(\tau_j(\pmb{p}+t_0))}\}.
\end{eqnarray}

It is noted that the CRB in (\ref{eq:crb}) with the EFIM in (\ref{eq:efim}) is increased by the presence of quantization as compared to the CRB of direct localization with unquantized signals derived in [8], [9], which is defined with the same EFIM $\pmb{J}(\pmb{p})$ in (\ref{eq:efim}) except for the different FIM $\pmb{\Psi}_j$ of each RU $j$ calculated with the nonzero elements for the non-orthogonal parameters as  
\begin{equation}\label{eq:psiIdeal} 
{[{\pmb{\Psi}_j}]_{p,q}}=\sum_{k,m,\zeta}\frac{2}{\sigma_j^2}\nabla_{[\tilde{\pmb{\theta}}_j]_p}f_{j,k,m}^\zeta(\pmb{\theta}_j)\nabla_{[\tilde{\pmb{\theta}}_j]_q}^Tf_{j,k,m}^\zeta(\pmb{\theta}_j).
\end{equation}

We can compare the FIM $\pmb{\Psi}_j$ of RU $j$ for quantized signals in (\ref{eq:psi}) and for unquantized signals in (\ref{eq:psiIdeal}) that appear in the EFIM $\pmb{J}(\pmb{p})$, in the low SNR regime, where the power of channel noise is much larger than that of the desired signal, i.e., $\sigma_j \gg |f_{j,k,m}(\pmb{\theta}_j)|$. For this purpose, we assume that all the RUs have the same fronthaul capacity constraints $B_j=B$ and that the channel noise power satisfies $\sigma_j^2=\sigma^2$ for $j \in \mathcal{N}_r$. Thus, we use the same uniform quantization with $L=2^{B/2M}$ quantization levels and quantization threshold $\{q_l(\Delta)\}$ for all RUs, for $1 \le l \le L$. Then, the FIM $\pmb{\Psi}_j$ of RU $j$ in (\ref{eq:psi}) for quantized case can be approximated as
\begin{equation}\label{eq:loss} 
{[{\pmb{\Psi}_j}]_{p,q}} = L_Q\sum_{k,m,\zeta}\frac{2}{\sigma^2}\nabla_{[\tilde{\pmb{\theta}}_j]_p}f_{j,k,m}^\zeta(\pmb{\theta})\nabla_{[\tilde{\pmb{\theta}}_j]_q}^Tf_{j,k,m}^\zeta(\pmb{\theta}),
\end{equation} 
where, by comparison to (\ref{eq:psiIdeal}) in the unquantized case, the factor $L_Q$ can be interpreted as the quantization loss of the CRB for quantized signals and is defined as
\begin{equation}\label{eq:LQ}
L_Q =\frac{1}{2\pi}\sum_{l=1}^{L}\frac{\left(e^{-\frac{q_{l}^2(\Delta)}{\sigma^2}}-e^{-\frac{q_{l-1}^2(\Delta)}{\sigma^2}}\right)^2}{\Phi\left(\frac{q_{l}(\Delta)}{\sigma/\sqrt{2}}\right)-\Phi\left(\frac{q_{l-1}(\Delta)}{\sigma/\sqrt{2}}\right)} \le 1.
\end{equation}
The factor $L_Q$ results in a decrease on the Fisher Information at each RU $j$ due to the operation over quantized signals, and we have that, from (\ref{eq:efim}), the EFIM $\pmb{J}(\pmb{p})$ of quantized case is equivalent to $L_Q$ times that of unquantized case. By the expression (\ref{eq:crb}) of the CRB, it can be seen that $L_Q$ directly evaluates the ratio between the CRBs of unquantized and quantized cases at low SNR, so we can approximately write   
\begin{equation}\label{eq:appLQ}
\hspace{1cm}L_Q \approx \frac{\text{CRB}^{\text{UQ}}}{\text{CRB}^{\text{Q}}},
\end{equation}  
where $\text{CRB}^{\text{UQ}}$ and $\text{CRB}^{\text{Q}}$ are the actual CRBs of direct localization in unquantized and quantized cases with the EFIM $\pmb{J}(\pmb{p})$ being given by (\ref{eq:psiIdeal}) and (\ref{eq:psi}), respectively. The approximation (\ref{eq:appLQ}) will be validated via numerical results in the next section. 

As a final remark, it is noted that, when the fronthaul capacity $B_j$ for $j \in \mathcal{N}_r$ becomes arbitrarily large, namely the step size of quantizer $\Delta_j$ goes to zero and the quantization level $L_j$ approaches to infinity, the CRB for quantized and unquantized signal models, $\text{CRB}^{\text{UQ}}$ and $\text{CRB}^{\text{Q}}$, can be seen to coincide. This can be proved by the fact that the FIM $\pmb{\Psi}_j$ of each RU $j \in \mathcal{N}_r$ in (\ref{eq:psi}) approaches to (\ref{eq:psiIdeal}), as $B_j$ turns to infinity (see Appendix \ref{app:proof}).  
\section{Numerical Results}\label{sec:num}
 
In this section, we evaluate the performance of direct localization with the aim of assessing the impact of quantization. For the indirect localization, as discussed in Section \ref{sec:InDi}, the source is located by ML based on TOA and AOA estimates both of which are obtained by means of the multiple signal classification (MUSIC) algorithm [5], [6], as in [8], but other super-resolution techniques are also possible. We consider a network with $N_r=4$ RUs placed at the vertices of a square area with side of length $D=4$ km, while the source is uniformly and randomly located within a $3\times3$ km$^2$ area of $A$ centered in the entire region. Each RU is assumed to be equipped with an $M=8$ element ULA with antenna separation $\Delta_A=\lambda/2$. The sampled waveform is $s(n)=\text{sinc}(t/T_s)/\sqrt{N_s}|_{t=nT_s}$ with bandwidth $W=1/(2T_s)=200$ kHz as in GSM, so that the sampling period is $T_s=2.5$ $\mu$s, and each location estimation is based on $N_s=8$ samples. We assume that the channel coefficients $b_j$ are independent and Rician, with Rician factor $K=20$ dB. Also, we assume sensors with equal noise variance. The dynamic range of the uniform quantizer applied at each RU $j$ to all antennas is set based on Monte Carlo experiments guaranteeing that the received signals are included in the peak-to-peak interval with the probability of $95$\% given conditions.

\begin{figure}[t]
\begin{center}
\includegraphics[width=10cm, height=8.5cm]{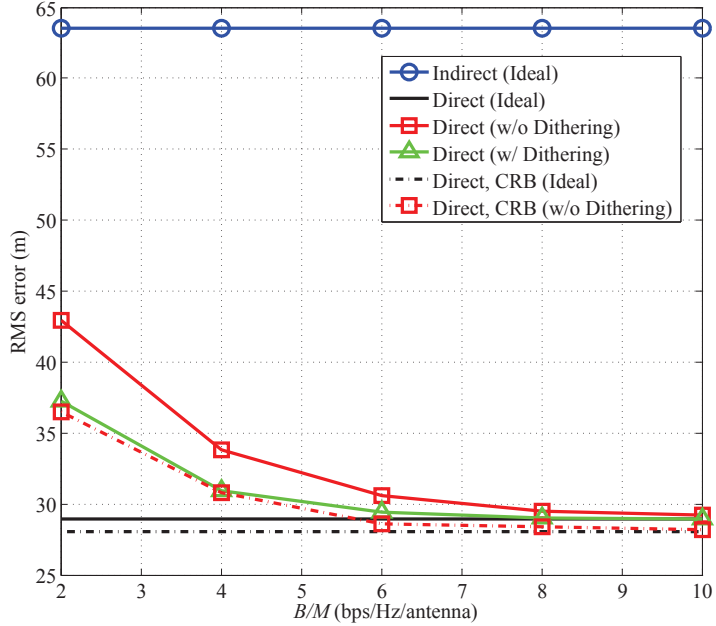}
\caption{RMS error versus fronthaul capacity $B/M$ bps/Hz/antenna for $\text{SNR (per antenna)}=0$ dB ($N_r=4$, $M=8$, $N_s=8$ and $T_s=2.5$ $\mu$s).} \label{fig:B0}
\end{center}
\end{figure} 

\begin{figure}[t]
\begin{center}
\includegraphics[width=10cm, height=8.5cm]{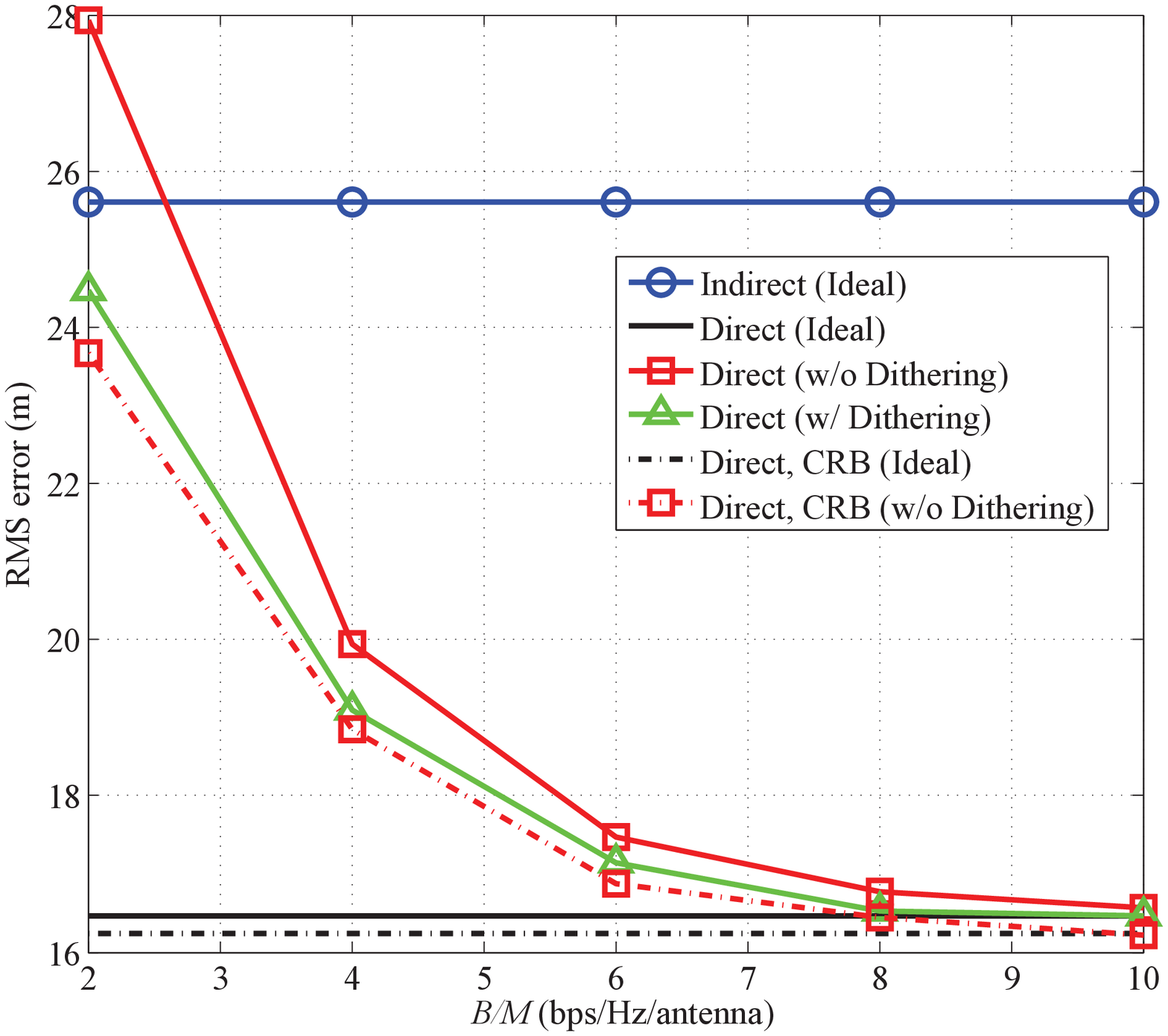}
\caption{RMS error versus fronthaul capacity $B/M$ bps/Hz/antenna for $\text{SNR (per antenna)}=5$ dB ($N_r=4$, $M=8$, $N_s=8$ and $T_s=2.5$ $\mu$s).} \label{fig:B5}
\end{center}
\end{figure}
We impose the fronthaul capacity constraints as $B_1=B_2=B_3=B$ and $B_4=2B$, and we set $q_{t_0}=1$ in the direct localization scheme for estimating $t_0$ (see Section \ref{sec:Di}). We evaluate the root mean squared (RMS) error as performance metric defined as $\sqrt{\sum_{n=1}^N\rho_n/N}$, where $\rho_n=\|\hat{\pmb{p}}_n-\pmb{p}\|^2$ with $N$ being the number of experiments and $\hat{\pmb{p}}_n$ being the estimated source location of the $n$th experiments. For reference, we also present the performance of direct localization with ideal fronthaul links and the CRBs for quantized and unquantized signals derived in Section \ref{sec:CRB}. 

Fig. \ref{fig:B0} and Fig. \ref{fig:B5} show the RMS error versus the fronthaul capacity constraint $B/M$ (bps/Hz /antenna) when SNR per antenna is $0$ dB and $5$ dB, respectively. First, we observe that, when the SNR is sufficiently large, as in Fig. \ref{fig:B5} (SNR$=5$ dB), indirect localization may outperform direct localization if the fronthaul capacity is small enough. This is due to the distortion caused by fronthaul quantization when the quantization is coarse. Note that this is not the case at lower SNR, where at SNR$=0$ dB, the signal degradation due to the additive noise overwhelms the loss due to quantization. Furthermore, as long as the fronthaul capacity is large enough, as expected, direct localization has the potential to significantly outperform indirect localization, with additional marginal gains achievable via dithering. Also, the gain of dithering is more relevant at lower fronthaul capacity. For instance, for SNR$=0$ dB, the gain of dithering (reduction in localization RMS) increases from $8.5$\% at $B/M=4$ to $13.3$\% at $B/M=2$. Finally, the CRB, as well as the performance with ideal fronthaul, is approached by the direct scheme as the fronthaul capacity and the SNR increase.
\begin{figure}[t]
\begin{center}
\includegraphics[width=10cm, height=8.5cm]{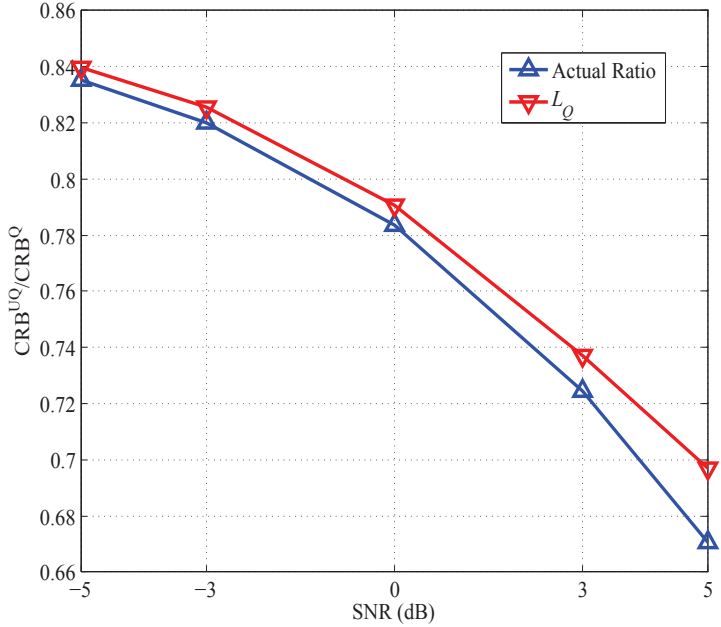}
\caption{Ratio between CRBs for unquantized and quantized observations versus SNR (per antenna) (dB) for $B/M=4$ bps/Hz/antenna ($N_r=4$, $M=8$, $N_s=8$ and $T_s=2.5$ $\mu$s).} \label{fig:LQ_SNR}
\end{center}
\end{figure} 

Next, we turn to the evaluation of the performance loss of direct localization due to quantization by means of the analysis in Section \ref{sec:CRB}. Specifically, we compare the quantization loss $L_Q$ in (\ref{eq:LQ}) to the actual ratio between the CRBs of the unquantized and quantized cases, i.e., $\text{CRB}^{\text{UQ}}/\text{CRB}^{\text{Q}}$, obtained by numerical simulations using the same parameters as discussed above, except that we impose an equal fronthaul capacity constraint $B_j=B$, and use the same uniform quantizer for all RU $j \in \mathcal{N}_r$.

Fig. \ref{fig:LQ_SNR} shows the discussed ratio between CRBs versus SNR per antenna for $B/M=4$ bps/Hz/antenna. It is seen that, the factor $L_Q$ provides a close approximation of the actual performance loss, as expected in (\ref{eq:appLQ}), not only in the low-SNR regime in which it was derived. In particular, the analytical loss (\ref{eq:LQ}) properly quantifies the quantization loss due to quantization, which becomes significant at the higher SNR.                
\section{Concluding Remarks}\label{sec:con}
The C-RAN architecture, with its centralized baseband processing at a control unit (CU), enables the implementation of direct, or one-step, localization. While direct localization outperforms the traditional indirect, or two-step, localization in the presence of ideal connections between the CU and the radio units (RUs), this may not be the case in the C-RAN due to the distortion caused by quantization on the fronthaul links. In this paper, we have studied the performance of direct and indirect localization in the C-RAN by means of the calculation of the CRB as well as quantization effect on the localization accuracy via analytical and numerical results. Interesting open problems concern the study of the optimal scalar quantization design for the RUs to maximize the localization accuracy.
\section{Acknowledgments}\label{sec:ack}
This work has been supported by the National GNSS Research Center program of Defense Acquisition Program Administration and Agency for Defense Development. 
{\parindent0pt
\parskip8pt
\section{References}
[1] Weiss, A. J.: 'Direct position determination of narrowband radio frequency transmitters', IEEE Signal Processing Letters, 2004, \textbf{11}, (5), pp. 513--516

[2] Jeong, S., Simeone, O., Haimovich, A.,~\textit{et al}.: 'Beamforming design for joint localization and data transmission in distributed antenna system', IEEE Transactions on Vehicular Technology, 2015, \textbf{64}, (1), pp. 62--76

[3] Jeong, S., Simeone, O., Haimovich, A.,~\textit{et al}.: 'Optimal fronthaul quantization for cloud radio positioning', IEEE Transactions on Vehicular Technology, 2015, \textbf{PP}, (99), pp. 1--7

[4] Gezici, S., Tian, Z., Giannakis, G. B.,~\textit{et al}.: 'Localization via ultra-wideband radios: a look at positioning aspects for future sensor networks', IEEE Signal Processing Magazine, 2005, \textbf{22}, (4), pp. 70--84

[5] Li, X., Pahlavan, K.: 'Super-resolution {TOA} estimation with diversity for indoor geolocation', IEEE Transactions on Wireless Communications, 2004, \textbf{3}, (1), pp. 224--234

[6] Schmidt, R. O.: 'Multiple emitter location and signal parameter estimation', IEEE Transactions on Antennas and Propagation, 1986, \textbf{34}, (3), pp. 276--280

[7] Bar-Shalom, O., Weiss. A. J.: 'Transponder-aided single platform geolocation', IEEE Transactions on Signal Processing, 2013, \textbf{61}, (5), pp. 1239--1248

[8] Weiss, A. J., Amar, A.: 'Direct position determination of multiple radio signals', Eurasip Journal Applied Signal Processing, 2005, \textbf{2005}, (1), pp. 37--49

[9] Amar, A., Weiss, A. J.: 'Direct position determination in the presence of model errors --known waveforms', Digital Signal Processing, 2006, \textbf{16}, (1), pp. 52--83

[10] Wax, M., Kailath, T.: 'Optimum localization of multiple sources by passive arrays', IEEE Transactions on Acoustics, Speech and Signal Processing, 1983, \textbf{31}, (5), pp. 1210--1217

[11] Wax, M., Kailath, T.: 'Decentralized processing in sensor arrays', IEEE Transactions on Acoustics, Speech, and Signal Processing, 1985, \textbf{33}, (5), pp. 1123--1129

[12] Niu, R., Varshney, P. K.: 'Target location estimation in sensor networks with quantized data', IEEE Transactions on Signal Processing, 2006, \textbf{54}, (12), pp. 4519--4528

[13] Hateld, D. N., 'A report on technical and operational issues impacting the provision of wireless enhanced 911 services' (Federal Communications Commission, Washington, D.C., 2002), pp. 1--54. Available at http://www.emergentcomm.\\com/documents/Supporting\_Documents/hatfield\_report1.pdf,  

[14] 3GPP TS 36.455, 'E-UTRA: LTE positioning protocol A (LPPa) v12.2.0', 2015
 
[15] Chen, K., Duan, R., 'C-RAN: The road towards green RAN' (China Mobile Research Institute, White Paper v2.5, 2011), http://labs.chinamobile.com/cran/wp-content/uploads/CRAN\_white\\ \_paper\_v2\_5\_EN.pdf, pp. 1--48

[16] Schuchman, L.: 'Dither signals and their effect on quantization noise', IEEE Transactions on Communication Technology, 1964, \textbf{12}, (4), pp. 162--165 

[17] Gray, R. M., Stockham, T. G.: 'Dithered quantizers', IEEE Transactions on Information Theory, 1993, \textbf{39}, (3), pp. 805--812

[18] Kay, S. M.: 'Fundamentals of signal processing-estimation theory' (Prentice Hall, NJ, 1993)

[19] CPRI: 'CPRI specification v6.1' (2014), http://www.cpri.info/jp/spec.html

[20] Cover, T. M., Thomas, J. A.: 'Element of information theory' (John Wiley \& Sons, NJ, 2006)

[21] Zhang, F.: 'The Schur complement and its applications' (Springer, NY, 2005), vol. 4.

[22] Zhang, H., Shen, Y., Chai, L., ~\textit{et al}.: 'State dimension reduction and analysis of quantized estimation systems', Signal Processing, 2014, \textbf{105}, pp. 363--375

[23] Olver, F. W. J.: 'NIST handbook of mathematical functions' (Cambridge University Press, NY, 2010)  

\section{Appendix}
\subsection{Calculation of EFIM for Direct Localization based on Quantized Signals}\label{app:efim}
In this appendix, we calculate the EFIM $\pmb{J}(\pmb{p})$ of direct localization in (\ref{eq:efim}). Similar to [2], [3], since the source is localizable, the mapping of $\pmb{\theta}=[\pmb{p}$$^T\,\,\pmb{b}_1^T \cdots\pmb{b}_{N_r}^T]^T$ to $\tilde{\pmb{\theta}}=[\tilde{\pmb{\theta}}_1^T \cdots \tilde{\pmb{\theta}}_{N_r}^T]^T$ with $\tilde{\pmb{\theta}}_j = [\tau_j(\pmb{p})\,\,\phi_j(\pmb{p})\,\,\pmb{b}_j^T]^T$ is a bijection. Then, we can have the relationship $\pmb{J}(\pmb{\theta})=\pmb{T}\pmb{J}(\tilde{\pmb{\theta}})\pmb{T}^T$, where $\pmb{T}=\partial \tilde{\pmb{\theta}}/\partial \pmb{\theta}=[\pmb{U}_1 \cdots \pmb{U}_{N_r};\,\,\pmb{V}_1 \cdots \pmb{V}_{N_r}]$ is the Jacobian matrix; and $\pmb{J}(\tilde{\pmb{\theta}})=-E_{\hat{\pmb{R}}}[\nabla_{\tilde{\pmb{\theta}}}\nabla_{\tilde{\pmb{\theta}}}^TL(\pmb{\theta})]$, with $\pmb{U}_j$ defined in (\ref{eq:efim}) and $\pmb{V}_j\in \Re^{2N_r\times 4}$ having all zeros except for $[\pmb{V}_j]_{2(j-1)+1:2j, 3:4}=\pmb{I}_2$. Referring [12], $\pmb{J}(\tilde{\pmb{\theta}})$ can be calculated as $\pmb{J}(\tilde{\pmb{\theta}})=\text{diag}\{\pmb{\Psi}_1, \dots,\pmb{\Psi}_{N_r}\}$, and using the relationship $\pmb{J}(\pmb{\theta})=\pmb{T}\pmb{J}(\tilde{\pmb{\theta}})\pmb{T}^T$, $\pmb{J}(\pmb{\theta})$ can be written as $\pmb{J}(\pmb{\theta})=[\pmb{X} \,\, \pmb{Y}; \pmb{Y}^T \,\, \pmb{Z}]$, where $\pmb{X}$, $\pmb{Y}$ and $\pmb{Z}$ are given in (\ref{eq:efim}). Finally, by applying the Schur complement (see, e.g., [21]), we can have the EFIM $\pmb{J}(\pmb{p})$ in (\ref{eq:efim}). 
\subsection{Proof of Asymptotic Convergence of CRB for Large Fronthaul Capacity}\label{app:proof}
Here, we show that the matrix $\pmb{\Psi}_j$ (\ref{eq:psi}) that appears in the CRB for quantized observations approaches the counterpart quantity (\ref{eq:psiIdeal}) for the case with unquantized observations as the fronthaul capacity $B_j$ for $j \in \mathcal{N}_r$ goes to infinity. This allows us to conclude, as desired, that the CRB with quantized observations tends to the CRB with unlimited fronthaul as $B_j$ tends to infinity. To this end, we first prove the limit  
\begin{eqnarray}\label{eq:Conv}
&&\hspace{0cm}\lim_{\Delta_j \to 0, L_j \to \infty}\sum_{l=1}^{L_j}\frac{\left(\Gamma_{j,k,m,l}^\zeta-\Gamma_{j,k,m,l-1}^\zeta\right)^2}{P_{j,l}(Q_j([\pmb{R}_j^\zeta(k)]_m); \pmb{\theta}_j)} \to 2\pi, 
\end{eqnarray}
for $\forall j,k,m,\zeta$. For simplicity of the notation, we will omit the subscript $k,m$ and superscript $\zeta$ in the following. The proof is similar to the proof of Theorem 2 in Section 4.2 of [22]. Denote $\tilde{q}_{j,l}=q_{j,l}(\Delta_j)-f_{j,k,m}^\zeta(\pmb{\theta}_j)$, and $\tilde{\Delta}_{j,l}=\tilde{q}_{j,l}-\tilde{q}_{j,l-1}$. Also, we define the probability density function of the standard normal distribution as $\phi(x)=e^{-x^2/2}/\sqrt{2\pi}$. Noting that $\tilde{\Delta}_{j,l} \to 0$ when $\Delta_j \to 0$, we can calculate the limit (\ref{eq:Conv}) as 
\begin{eqnarray}\label{eq:reConv}
&&\hspace{-1cm}\lim_{\tilde{\Delta}_{j,l} \to 0, L_j \to \infty}\sum_{l=1}^{L_j}\frac{2\pi\left(\phi\left(\frac{\tilde{q}_{j,l}}{\sigma_j/\sqrt{2}}\right)-\phi\left(\frac{\tilde{q}_{j,l-1}}{\sigma_j/\sqrt{2}}\right)\right)^2}{\Phi\left(\frac{\tilde{q}_{j,l}}{\sigma_j/\sqrt{2}}\right)-\Phi\left(\frac{\tilde{q}_{j,l-1}}{\sigma_j/\sqrt{2}}\right)}\nonumber\\
&&\hspace{-1.5cm}=\sum_{l=1}^{\infty}\lim_{\tilde{\Delta}_{j,l} \to \infty}2\pi\left(\frac{\phi\left(\frac{\tilde{q}_{j,l}}{\sigma_j/\sqrt{2}}\right)-\phi\left(\frac{\tilde{q}_{j,l}-\tilde{\Delta}_{j,l}}{\sigma_j/\sqrt{2}}\right)}{\frac{\tilde{\Delta}_{j,l}}{\sigma_j/\sqrt{2}}}\right)^2\left(\frac{\Phi\left(\frac{\tilde{q}_{j,l}}{\sigma_j/\sqrt{2}}\right)-\Phi\left(\frac{\tilde{q}_{j,l}-\tilde{\Delta}_{j,l}}{\sigma_j/\sqrt{2}}\right)}{\frac{\tilde{\Delta}_{j,l}}{\sigma_j/\sqrt{2}}}\right)^{-1}\frac{\tilde{\Delta}_{j,l}}{\sigma_j/\sqrt{2}}\nonumber\\
&&\hspace{-1.5cm}=\sum_{l=1}^{\infty}\lim_{\tilde{\Delta}_{j,l} \to \infty}2\pi\left(\left.\frac{d}{dt}\phi(t)\right|_{t=\frac{\tilde{q}_{j,l}}{\sigma_j/\sqrt{2}}}\right)^2\left(\left.\frac{d}{dt}\Phi(t)\right|_{t=\frac{\tilde{q}_{j,l}}{\sigma_j/\sqrt{2}}}\right)^{-1}\frac{\tilde{\Delta}_{j,l}}{\sigma_j/\sqrt{2}}\nonumber\\
&&\hspace{-1.5cm}=\sum_{l=1}^{\infty}\lim_{\tilde{\Delta}_{j,l} \to \infty}\sqrt{2\pi}\left(\frac{\tilde{q}_{j,l}}{\sigma_j/\sqrt{2}}\right)^2e^{-\frac{\tilde{q}_{j,l}^2}{\sigma_j^2}}\frac{\tilde{\Delta}_{j,l}}{\sigma_j/\sqrt{2}}\nonumber\\
&&\hspace{-1.5cm}=\int_{\pmb{R}}\sqrt{2\pi}\left(\frac{\tilde{q}_{j,l}}{\sigma_j/\sqrt{2}}\right)^2e^{-\frac{\tilde{q}_{j,l}^2}{\sigma_j^2}}\frac{d\tilde{q}_{j,l}}{\sigma_j/\sqrt{2}}\nonumber\\
&&\hspace{-1.5cm}=\int_{\pmb{R}}\sqrt{2\pi}r^2e^{-\frac{r^2}{2}}dr \to 2\pi.
\end{eqnarray}
The last equation can be obtained by denoting $r=\tilde{q}_{j,l}/(\sigma_j/\sqrt{2})$ and by an Euler integral of Gamma function $\int_{\pmb{R}}r^2e^{-r^2/2}/\sqrt{2\pi}dr=1$ [23]. Given the limit (\ref{eq:Conv}), we now have that the limit 
\begin{eqnarray}\label{eq:th1}
&&\hspace{-2cm}\lim_{\Delta_j \to 0, L_j \to \infty}\sum_{k,m,\zeta}\sum_{l=1}^{L_j}\frac{\left(\Gamma_{j,k,m,l}^\zeta-\Gamma_{j,k,m,l-1}^\zeta\right)^2\nabla_{[\tilde{\pmb{\theta}}_j]_p}f_{j,k,m}^\zeta(\pmb{\theta}_j)\nabla_{[\tilde{\pmb{\theta}}_j]_q}^Tf_{j,k,m}^\zeta(\pmb{\theta}_j)}{\pi\sigma_j^2 P_{j,l}(Q_j([\pmb{R}_j^\zeta(k)]_m); \pmb{\theta}_j)}\nonumber\\
&&\hspace{5cm}= \sum_{k,m,\zeta}\frac{2}{\sigma_j^2}\nabla_{[\tilde{\pmb{\theta}}_j]_p}f_{j,k,m}^\zeta(\pmb{\theta}_j)\nabla_{[\tilde{\pmb{\theta}}_j]_q}^Tf_{j,k,m}^\zeta(\pmb{\theta}_j)
\end{eqnarray} 
holds, which concludes the proof since (\ref{eq:th1}) shows that the EFIM for the case with quantized observations tends to that of the case with unquantized observations, as the fronthaul capacity $B_j$ for $j \in \mathcal{N}_r$ goes to infinity.

\end{document}